\documentstyle[prb,aps,twocolumn]{revtex} 
\begin{document} 
\draft
\title{Linear Field Dependence of the Normal-State In-Plane
Magnetoresistance of ${\text {Sr}}_2{\text {RuO}}_4$} 

\twocolumn[
\hsize\textwidth\columnwidth\hsize\csname@twocolumnfalse\endcsname

\author{R. Jin, Y. Liu} 
\address{Department of Physics, The Pennsylvania State
University, University Park, PA 16802} 
\author{F. Lichtenberg}
\address{ Center for Electronic Correlations and Magnetism, Augsburg
University, D-86135 Augsburg, Germany} 

\date{\today} 
\maketitle

\begin{abstract}

The transverse and longitudinal in-plane magnetoresistances in the
normal state of superconducting ${\text {Sr}}_2{\text {RuO}}_4$ single 
crystals have been measured.   At low
temperatures, both of them were found to be positive 
with a linear magnetic-field dependence above a threshold 
field, a result not expected from electronic band theory.  We
argue that such behavior is a manifestation of a novel coherent
state characterized by a spin pseudo gap in the quasi-particle
excitation spectrum in ${\text {Sr}}_2{\text {RuO}}_4$.

\end{abstract}

]


\narrowtext

Since the discovery of superconductivity in 
${\text {Sr}}_2{\text {RuO}}_4$,\cite{1} 
isostructural with high-$T_{\text c}$ cuprate
${\text {(La,Sr)}}_2{\text {CuO}}_4$,\cite{1,2} 
this material has emerged as a new focus of superconductivity
research.\cite{3}  There has been a growing body of experimental
evidence for a spin triplet (odd-parity) pairing in 
${\text {Sr}}_2{\text {RuO}}_4$,\cite{4}
including results obtained in muon spin relaxation,\cite{5} NMR
Knight shift, \cite{6} electrical transport,\cite{7} proximity
effect,\cite{8} and specific heat\cite{9} measurements.  In the normal
state, a highly anisotropic Fermi liquid behavior, characterized by
a ${T}^2$-dependence in the in- and out-of-plane resistivity 
($\rho _{ab}$ and $\rho _c$, respectively)\cite{3} 
and a cylindrical Fermi surface,\cite{10} was found for 
${\text {Sr}}_2{\text {RuO}}_4$ at
low temperatures. The enhancements of electronic specific heat and
spin susceptibility are very similar to those found in 
${}^3{\text {He}}$,\cite{3,4} suggesting that 
${\text {Sr}}_2{\text {RuO}}_4$ is
a strongly correlated system.  At higher temperatures, however, the
situation is not very clear.  While $\rho _c$ undergoes a
nonmetallic (d$\rho _c/dT<0$) to metallic (d$\rho _c/dT>0$)
crossover around 130K,\cite{1,2} a similar feature is not seen in 
$\rho _{ab}$, a phenomenon yet to be fully understood.

Chemically, 
${\text {Sr}}_2{\text {RuO}}_4$ 
is positioned in the vicinity of both ferromagnetic and
antiferromagnetic orderings. In the Ruddlesden-Popper 
homologous series 
${\text {Sr}}_{n+1}{\text {Ru}}_n{\text O}_{3n+1}$, 
as the single-layer (n $=1$) member,
${\text {Sr}}_2{\text {RuO}}_4$ 
is a paramagnetic metal with no apparent local moment.\cite{11}  
The infinite-layer  (${\text n}=\infty $) member, 
${\text {SrRuO}}_3$, 
is a ferromagnetic metal.\cite{12}  The n = 2 member of the series,
${\text {Sr}}_3{\text {Ru}}_2{\text O}_7$, 
which has been reported to be ferromagnetic by one group\cite{13}
and antiferromagnetic by others,\cite{14,15} should be magnetic as
well. On the other hand, if ${\text {Sr}}^{2+}$ ions in 
${\text {Sr}}_2{\text {RuO}}_4$ are replaced by 
${\text {Ca}}^{2+}$, the resultant compound 
${\text {Ca}}_2{\text {RuO}}_4$, 
which is isostructural and isoelectronic to
${\text {Sr}}_2{\text {RuO}}_4$, is an antiferromagnetic
insulator.\cite{16,17} Being in the vicinity of these very 
different behaviours, 
${\text {Sr}}_2{\text {RuO}}_4$ is likely to exhibit
unusual electronic and magnetic properties.

The normal-state properties of 
${\text {Sr}}_2{\text {RuO}}_4$ 
can be probed by magnetoresistance (MR) 
measurements.\cite{18} The normal-state MR of 
${\text {Sr}}_2{\text {RuO}}_4$ 
was first measured at 20mK in the context of determining the 
Fermi surface of this material through the Shubnikhov-de Haas
oscillations.\cite{19} However, no field dependence of MR was
identified in that work. More recently, Hussey {\it et al.} have
measured both {\it c}-axis and in-plane normal-state MR of 
${\text {Sr}}_2{\text {RuO}}_4$, focusing primarily on the 
{\it c}-axis MR
($\Delta \rho _c/\rho _c = (\rho _c(H)-\rho_c(0))/\rho _c(0)$) 
at relatively high temperatures.\cite{20} 
The in-plane MR 
($\Delta \rho _{ab}/\rho _{ab}=(\rho _{ab}(H)-\rho_{ab}(0))/\rho
_{ab}(0)$)
was measured only in transverse configuration
 for one sample, without a
detailed analysis of its field dependence over the entire temperature
range.
Longitudinal in-plane MR results were not available in literature
prior to the present study.  In this paper, we report
results of our systematic study of the normal-state in-plane MR for
${\text {Sr}}_2{\text {RuO}}_4$ in both transverse and longitudinal 
configurations. Our work has revealed previously
unknown features in  in-plane MR, 
strongly suggesting a novel behavior in the normal state of 
${\text {Sr}}_2{\text {RuO}}_4$.

Single crystals used in this study were grown by a floating-zone
method with details described elsewhere.\cite{2} Resistance
measurements show a superconducting transition temperature ($T_c$)
of 0.84K. MR measurements were carried out in a ${}^3{\text {He}}$
cryostat.  The temperature was measured using a Lakeshore Cernox
1030 thermometer with relative temperature corrections, due to
magnetic field, typically 0.15\% at 4.2 K and -0.023\% at 77.8K and 8.0T.
\cite{21b} 
For in-plane MR and Hall
measurements, we used four rectangular shaped single crystals with
dimensions around 
$1.2\times 1.0\times 0.01$, $0.9\times 0.4\times 0.08$, 
$0.8\times 0.2\times 0.07$, and 
$0.6\times 0.3\times 0.05{\text {mm}}^3$, 
respectively. For each sample, two current contacts
covering the opposite ends and four voltage contacts on the two
sides of the crystal were prepared.  All ${\text {RuO}}_2$ layers
were electrically shorted along the {\it c}-axis to ensure a
homogeneous current distribution.  For transverse and longitudinal
MR measurements, the magnetic field $H$ was applied perpendicular
and parallel to the injected current $I$, respectively.  In order to
exclude the Hall contribution to MR, only the symmetric part
of $\Delta \rho _{ab}(H)=\rho _{ab}(H) - \rho _{ab}(0)$ 
under field reversal was included.  For in-plane Hall measurements,
the magnetic field was applied parallel to $c$-axis with a current bias
applied along the $ab$-plane.  The Hall voltage $V_{\text H}$, which
contains only the asymmetric contributions under the field reversal,
was found to vary linearly with $H$ up to 5T.  By fitting 
$V_{\text H}(H)$
data in the linear regime using 
$V_{\text H}=R_{\text H}\cdot H\cdot I/d$ 
($d$ is the thickness of the sample), the in-plane Hall coefficient
$R_{\text H}$ was obtained.

The transverse in-plane MR, $\Delta \rho _{ab}^{\perp }/\rho _{ab}$ 
 ($H \perp I $), between 0 and 7.3T at
various temperatures is shown in Fig. \ref{Fig1}.  Similar results
have been obtained in separate samples. 
$\Delta \rho _{ab}^{\perp }/\rho _{ab}$ 
is seen to be positive, growing rapidly in magnitude
with decreasing $T$.  At low fields, 
$\Delta \rho _{ab}^{\perp }/\rho _{ab}$ 
at a fixed temperature is very small, which can be
described by the sum of a linear and a quadratic term. When $H$
exceeds a threshold value $H_0$, 
$\Delta \rho _{ab}^{\perp }/\rho _{ab}=K_{\text S}(H - H_0)$, 
where $K_{\text S}$ and $H_0$ are functions of
temperature. The solid lines in Fig. \ref{Fig1} represent fits to
the data above $H_0$ using this form.  The temperature dependence of
$H_0$ is shown in the inset of Fig. \ref{Fig1}.  We note that in Ref. 20, 
results on $\Delta \rho _{ab}^{\perp }/\rho _{ab}$, plotted
against $H^2$, were presented.  While 
$\Delta \rho _{ab}^{\perp }/\rho _{ab}$ 
might be quadratic in $H$ at the two highest temperatures shown $($56K
and 82K$)$, a deviation from this behavior can be seen at all other
temperatures.  (We have re-plotted the low temperature data against
$H$ and found both the magnitude and the field dependence of 
$\Delta \rho _{ab}^{\perp }/\rho _{ab}$ 
given in Ref. 20 are in agreement with those obtained in the present
study at overlapping temperatures.)

Within the band theory for solids, MR is proportional to 
$(H/\rho (0))^2$ in the low-field limit.\cite{18} Linear MR can be
expected only in some special circumstances.  For example, a square
Fermi surface can lead to linear MR due to the presence of the sharp
corners.\cite{18} Local-density approximation calculations have
shown that the Fermi surface of ${\text {Sr}}_2{\text {RuO}}_4$
consists of three roughly cylindrical-like sheets with no such sharp
corners,\cite{21} thus excluding this from being the origin of the
observed linear MR.  For systems with multi-band electronic
structure involving two types of charge carriers, the field
dependence of MR can be written as 
$\Delta \rho /\rho (0)={\text a}H^2/({\text b}+{\text c}H^2)$ 
where a, b, and c are
positive, field-independent quantities determined by the relaxation
rates of each type of carriers.\cite{22} It is clear that this
expression will not lead to
$\Delta \rho /\rho (0)\sim H$ 
as observed experimentally.  In some elemental metals (such as K, In,
and Al), linear MR has been found and attributed to the boundary and
disorder effects.\cite{18} Since superconductivity in 
${\text {Sr}}_2{\text {RuO}}_4$ is extremely
sensitive to the amount of disorder,\cite{7} the linear 
$\Delta \rho _{ab}^{\perp }/\rho _{ab}$, found reproducibly in
superconducting 
${\text {Sr}}_2{\text {RuO}}_4$ with 
$\rho _{ab}(0)\approx 1\mu \Omega $cm, is
unlikely due to effects of the impurities and/or structural defects.

Linear MR was observed previously in heavy fermion 
${\text {CeCu}}_6$ at very low temperatures.\cite{23}  In another
heavy fermion compound, ${\text {UPt}}_3$, the MR was found \cite{24} 
to be positive and proportional to $H^{1.25}$. The observed 
linear MR was interpreted as being a consequence
of the opening of a pseudo gap in a coherent state formed at low
temperatures.\cite{25} In this picture, the pseudo gap is 
suppressed by the magnetic
field, resulting in enhanced scattering rates and therefore positive MR. 
NMR studies of ${\text {Sr}}_2{\text {RuO}}_4$ provided the first
hint that a coherent normal state may also be present 
in this material at low
temperatures . The emergence of such a state is signaled by a sharp
change of slope in $1/T_1$ vs. temperature curve for both
${}^{101}{\text {Ru}}$ and planar ${}^{17}{\text O}$ sites around
80K.\cite{27}  Our in-plane MR results also indicate that 
this temperature is special for 
${\text {Sr}}_2{\text {RuO}}_4$.  
As shown in Fig. \ref{Fig2}, a sign change from
positive to negative in 
$\Delta \rho _{ab}^{\perp }/\rho _{ab}$ 
is seen between 70-80 K, similar to what has been observed in  
{\it c}-axis MR. \cite{20} It should be noted that the range of the 
data scattering in
$\Delta \rho _{ab}^{\perp }/\rho _{ab}$
at 7.3 T is around 0.004\%.  The corrections in temperature
due to the MR of the thermometer is around
-0.016\% at 87 K and 8.0 T, \cite {21b}
which cannot lead to the observed sign change in
 $\Delta \rho _{ab}^{\perp }/\rho _{ab}$. 
While the physical origin of 
the sign reversal in MR has not been 
understood at the present time, it is not inconceivable that this 
is associated with the emergence of a coherent normal state in 
${\text {Sr}}_2{\text {RuO}}_4$ at low temperatures.  As will be
argued later, such a state
may be accompanied by the opening of a pseudo gap in the 
quasi-particle excitation spectrum.  Similar to ${\text {CeCu}}_6$,
an applied magnetic field suppresses the pseudo gap,  leading to linear
in-plane MR for ${\text {Sr}}_2{\text {RuO}}_4$.

The in-plane Hall coefficient $R_{\text H}$ of 
${\text {Sr}}_2{\text {RuO}}_4$,
should be subject to the same scattering processes as that for the
in-plane MR. An interesting question is whether $R_{\text H}$ will 
reflect the presence of a pseudo gap in the quasi-particle excitation
spectrum. The in-plane $R_{\text H}$ was measured for one of our samples
in
the same magnetic field and current configuration as that for the
transverse in-plane MR. The results are shown in Fig. \ref{Fig3}.
The Hall coefficient is negative at high temperatures, changing its
sign around 130K, which is roughly where $\rho _c$ undergoes a
nonmetallic-metallic crossover. It reaches a maximum between 70-80K
before changing its sign again around 30K.  Prior to the present
work, the Hall coefficient $R_{\text H}$ of 
${\text {Sr}}_2{\text {RuO}}_4$
has been measured on crystals prepared by different
groups.\cite{27,28} Our results agree well with those published
earlier, indicating that features shown in $R_{\text H}(T)$ are intrinsic.
In Ref. 19, an isotropic-$\ell $ approximation within the multi-band
picture of 
${\text {Sr}}_2{\text {RuO}}_4$ was used to calculate the
low temperature limit ($T<1$K) of $R_{\text H}$ with band 
parameters obtained from quantum oscillations.  The calculated
result agrees with the experimental value. On the other hand, a
manybody-theory-based calculation of $R_{\text H}(T)$ has also been 
carried out,\cite{29} yielding a result 
that is consistent with experimental observations.  This suggests that an
account for behaviors in $R_{\text H}$ beyond band theory is still
possible,
especially for $T>1$K where the isotropic-$\ell $ approximation did
not seem to work.\cite{19} It is interesting to note that the
characteristic
temperature where $R_{\text H}$ shows a maximum coincides with that below
which we believe the pseudo gap emerges.  Whether the maximum in
$R_{\text H}$ is another signature for the opening of a pseudo gap in
${\text {Sr}}_2{\text {RuO}}_4$, and how strongly the presence of such a
gap will affect $R_{\text H}$, are issues to be resolved.

In order to infer the relative contribution of the orbital and spin
motions to MR and the physical origin of this pseudo gap, we have
measured the longitudinal in-plane MR of 
${\text {Sr}}_2{\text {RuO}}_4$.  As shown in Fig. \ref{Fig4}, 
the longitudinal in-plane MR
$\Delta \rho _{ab}^{\parallel }/\rho _{ab}$ 
also revealed a linear behavior above $H_0$, which increases with
increasing
temperature (see the inset of Fig. \ref{Fig4}).  
More striking is that the magnitude of 
$\Delta \rho _{ab}^{\parallel }/\rho _{ab}$ is greater than that of
$\Delta \rho _{ab}^{\perp }/\rho _{ab}$ below
approximately 10K, indicating that the spin
contribution to the in-plane MR is important. Therefore, the pseudo
gap may be of spin origin. Results
obtained in NMR ${}^{17}$O Knight shift measurements in 
${\text {Sr}}_2{\text {RuO}}_4$\cite{26,30} point to the same
direction. When the ${}^{17}$O Knight shift results were decomposed
into spin susceptibilities of three Ru d-orbitals, 
$d_{xy}$, $d_{xz}$, and $d_{yz}$, 
a broad maximum around 40K was found. 
This is consistent with uniform susceptibility results where a
broad peak was seen around the same temperature.\cite{11}  The
decrease of spin susceptibility at low temperatures may then be
taken as an indication of the opening of a pseudo gap in the spin
excitation spectrum.

It has been suggested theoretically that the spin fluctuations in
${\text {Sr}}_2{\text {RuO}}_4$ are predominantly
ferromagnetic.\cite{4,31} Experimentally, $1/T_1$ obtained from the
NMR measurements for both O(1) and Ru sites shows an identical
temperature dependence.\cite{26}  In general, $1/T_1$ can be
expressed by the sum of {\bf q}-dependent imaginary part of the
dynamical electron susceptibility with appropriate form factors,
which should have different {\bf q}-dependence for ${}^{101}$Ru and
planar ${}^{17}$O sites, leading to different temperature dependence
in their respective $1/T_1$. However, if the dynamical electron 
susceptibility has a
sharp peak around ${\bf q}=(0, 0)$, corresponding to ferromagnetic
fluctuations, then the above mentioned behavior in $1/T_1$ can be
explained.  On the other hand, this scenario will lead to negative
MR,\cite{18} which contradicts the experimental observation of
positive MR.

Results obtained in inelastic neutron scattering (INS) measurements
carried out recently on ${\text {Sr}}_2{\text {RuO}}_4$\cite{32}
have resolved this apparent contradiction between NMR and
magnetotransport results. In this experiment, 
incommensurate magnetic spin fluctuations located at 
${\bf q}_0=(\pm 0.6/a, \pm0.6/a, 0)$, 
originally predicted by band calculations,\cite{33}
were found at low temperatures. The intrinsic {\bf q}-width
for these incommensurate fluctuations was found to be narrow 
(${\Delta \bf q}=0.13\pm 0.02$\AA \enskip at 10K). 
The temperature dependence of $1/T_1$
observed in NMR can be reconstructed from the INS results,
confirming that the dominating spin fluctuations are incommensurate
rather than ferromagnetic as suggested in Ref. 26.  The energy
dependence of the imaginary part of the dynamic susceptibility at
${\bf q}_0$, which measures the dissipation of the spin
fluctuations, shows a gradual drop below 7meV, indicating the
presence of a gap in the spin excitation spectrum.  We note here
that this energy scale (7 meV) agrees well with the value of the pseudo
gap inferred from the temperature dependence of other physical
quantities such as $1/T_1$ and MR.

Physical insight into the nature of the normal state of 
${\text {Sr}}_2{\text {RuO}}_4$
can be obtained from the high-pressure experiments. 
It was shown that the superconducting transition temperature 
$T_c$ was suppressed by an applied pressure 
$p$.\cite{33b} 
By extrapolating results obtained up to 1.2 GPa, 
$T_c$ is expected to become zero around 
$p_c\approx $ 3 GPa. In normal state, 
$\rho _{ab}$ was found to vary non-monotonically with 
$p$ at low $T$.\cite{33c}  With increasing
$p$, $\rho _{ab}$  first increases, and then decreases as
$p$ was increased further above a threshold value, 
close to 3 GPa. At 8 GPa, 
$\rho _{ab}$ was found to show a 
$T^{4/3}$-dependence, characteristic of a 
two-dimensional electronic system close to a 
ferromagnetic quantum critical point.\cite{33d}
  Based on this, we propose the following phase diagram for
${\text {Sr}}_2{\text {RuO}}_4$, 
shown schematically in Fig.\ref{Fig5}. 
At $T$ = 0 K, a direct superconductor (SC)-to-ferromagnetic 
metal (FM) transition, tuned by pressure, occurs at 
$p=p_c\approx $ 3 GPa.  At low 
$T$ and low $p$,
${\text {Sr}}_2{\text {RuO}}_4$ is superconducting.  
A pseudo gap regime emerges above $T_c$,
 similar to the quantum spin disordered regime 
discussed in quantum critical phenomena.\cite{33e} 
As $p$ is further increased, the pseudo gap is suppressed, 
leading to an increase in $\rho _{ab}$, 
as observed experimentally.\cite{33c}  
When the system enters the fluctuating 
regime at a finite $T$,
 $\rho _{ab}$ decreases with the increasing 
$T$ since the system is driven towards a ferromagnetic state.  
We suggest that the magnetic field similarly 
suppresses the pseudo gap, resulting in a positive MR.  
In this picture, as $T$ increases, a larger threshold field 
($H_0$) is needed to overcome the thermal smearing 
so that linear MR can be observed, consistent 
with our experimental observations.

A pseudo gap has been observed in high-$T_c$ cuprate
superconductors.\cite{34}  However, the origin and physical 
consequences of pseudo gap in high-$T_c$ materials 
may be different from that in
${\text {Sr}}_2{\text {RuO}}_4$. 
The in-plane MR observed in 
${\text {(La,Sr)}}_2{\text {RuO}}_4$\cite{35} is positive, 
but not linear in
field dependence.  When the pseudo gap opens, the 
Hall coefficient was found to become strongly 
temperature-dependent.\cite{36}  A suppression in
resistivity was found in underdoped 
${\text {(La, Sr)}}_2{\text {RuO}}_4$ 
as the pseudo gap opened.\cite{35}  A similar feature has
not been observed in ${\text {Sr}}_2{\text {RuO}}_4$. As mentioned
above, the pseudo gap in 
${\text {Sr}}_2{\text {RuO}}_4$ 
is present only in a small portion of the
Fermi surface.  The magnitude of the gap is also small as compared
with that of the high-$T_c$ cuprates, resulting in unobservable
effects in $\rho _{ab}$ and $\rho _c$.

Finally we briefly mention our 
$\Delta \rho _{c}/\rho _{c}$
 results which will be discussed in detail in a future publication.  
As mentioned above, previously published work on 
$\Delta \rho _{c}/\rho _{c}$
 of ${\text {Sr}}_2{\text {RuO}}_4$ 
was focused on relatively high-temperature 
($>$22 K) behavior although data obtained at 3.6 K 
were also shown.\cite{20}  
The magnitude of our longitudinal $c$-axis MR, 
$\Delta \rho _{c}^{\parallel }/\rho _{c}$,
is comparable with that of the transverse MR,
$\Delta \rho _{c}^{\perp }/\rho _{c}$ . 
The difference, 
 $\Delta \rho _{c}^{\perp }/\rho _{c}$
- $\Delta \rho _{c}^{\parallel }/\rho _{c}$, 
the "pure" orbital MR, has very good 
$H^2$-dependence.  A sign reversal from positive at low 
$T$ to negative at high
$T$ was found in both 
$\Delta \rho _{c}^{\perp }/\rho _{c}$  and 
$\Delta \rho _{c}^{\parallel }/\rho _{c}$  around 75 K, 
consistent with previous observations.\cite{20}  
However, different from the previous study, 
we found that
$\Delta \rho _{c}(H)/\rho _{c}=\alpha H^2-\beta H^4$ (where $\alpha $ and 
$\beta $ are constants), actually describes well our
$\Delta \rho _{c}^{\perp }/\rho _{c}$ and
$\Delta \rho _{c}^{\parallel }/\rho _{c}$  data.\cite{36b}

In summary, we have studied the magnetic field dependence of in-plane 
MR of ${\text {Sr}}_2{\text {RuO}}_4$ up to 7.3T. The
linear positive transverse and longitudinal in-plane MR, not expected 
from band theory, has been found above a threshold field at low 
temperatures.  We argue that such behavior is a manifestation of a 
coherent pseudo gap state in 
${\text {Sr}}_2{\text {RuO}}_4$ 
formed at low temperatures. Theoretical input is needed to fully 
understand the experimental results.

We would like to acknowledge helpful discussions with P. Coleman, N.
Hussey, T. Imai, J. Jain, H. Kawano, Y. B. Kim, and G. Lonzarich,
technical assistance from Yu. Zadorozhny, and finally many important
contributions of D. Schlom to this project. This work is supported
in part in the US by NSF through grants DMR-9702661 and ECS-9705839
and by the BMBF (Project number 13N6918/1) in Germany.



\begin{figure} 
\caption{In-plane transverse MR, $\Delta \rho
_{ab}^{\perp }/\rho _{ab}$ ($H\perp {ab}$, $I\parallel ab$), for
${\text {Sr}}_2{\text {RuO}}_4$.  The solid lines are linear fits to
the data (see text). $H_0(T)$ is shown in the inset.} 
\label{Fig1}
\end{figure}

\begin{figure} 
\caption{In-plane transverse MR for ${\text {Sr}}_2{\text {RuO}}_4$
at temperatures close to 75K.} 
\label{Fig2} 
\end{figure}

\begin{figure} 
\caption{In-plane Hall coefficient $R_{\text H}(T)$ with a
broad peak around 70-80K.} 
\label{Fig3} 
\end{figure}

\begin{figure} 
\caption{In-plane longitudinal MR $\Delta \rho
_{ab}^{\parallel }/\rho _{ab}$ ($H\parallel {ab}$, $I\parallel ab$)
for ${\text {Sr}}_2{\text {RuO}}_4$.  The solid lines are linear
fits to the data.  $H_0(T)$ is shown in the inset. Note that $\Delta
\rho _{ab}^{\parallel }/\rho _{ab}$ is larger than $\Delta \rho
_{ab}^{\perp }/\rho _{ab}$ below 10K.} 
\label{Fig4} 
\end{figure}

\begin{figure} 
\caption{Schematic phase diagram
for ${\text {Sr}}_2{\text {RuO}}_4$,  where SC denotes 
the superconducting while FM the ferromagnetic metal phase.  
A zero-temperature SC-FM transition at 
$p=p_c\approx $ 3 GPa is indicated (see text).} 
\label{Fig5} 
\end{figure}


\begin{references}

\bibitem{1} Y. Maeno, {\it et al.}, Nature {\bf 372}, 532 (1994).

\bibitem{2} F. Lichtenberg, {\it et al.}, Appl. Phys. Lett. {\bf
60}, 1138 (1992).

\bibitem{3} Y. Maeno, Cze. J. Phys. {\bf 46}, Suppl. S6, 3097
(1996).

\bibitem{4} T.M. Rice and M. Sigrist, J. Phys.: Cond. Matt. {\bf 7},
L643 (1995).

\bibitem{5} G.M. Luke, {\it et al.}, Nature {\bf 394}, 558 (1998).

\bibitem{6} K. Ishida, {\it et al.}, Nature {\bf 396}, 658 (1998).

\bibitem{7} A.P. Mackenzie, {\it et al.}, Phys. Rev. Lett. {\bf 80},
161(1998).

\bibitem{8} R. Jin, {\it et al.}, Phys. Rev. B {\bf 59}, 4433
(1999).

\bibitem{9} S. Nissshizaki, {\it et al.}, J. Phys. Soc. Jpn. {\bf
67}, 560(1998).

\bibitem{10} A.P. Mackenzie, {\it et al.}, Phys. Rev. Lett. {\bf
76}, 3786 (1996).

\bibitem{11} Y. Maeno {\it et al.}, J. Phys. Soc. Japan. {\bf 66},
1405 (1997).

\bibitem{12} A. Callghan, C.W. Moeller, and R. Ward, Inorganic Chem.
{\bf 5}, 1572 (1966).

\bibitem{13} G. Cao, S. McCall and J.E. Crow, Phys. Rev. B {\bf 55},
R672 (1997).

\bibitem{14} R.J. Cava, {\it et al.}, J. of Solid State Chem. {\bf
116}, 141 (1995).

\bibitem{15} Y. Maeno, S. Nakatsuji, and S. Ikeda, Proc. Taniguchi
Symp. on Phys. and Chem. of Transition Metal Oxides, Springer Series
in Solid-State Sci., Vol. {\bf 125} (1998).

\bibitem{16} G. Cao, {\it et al.}, Phys. Rev. B {\bf 56}, R2916
(1997).

\bibitem{17} S. Nakatsuji, S. Ikeda, and Y. Maeno, J. Phys. Soc.
Jpn. {\bf 66}, 1868 (1997).

\bibitem{18} A.B. Pippard, {\it Magnetoresistance in Metals},
(Cambridge Univ. Press, 1989).

\bibitem{19} A.P. Mackenzie {\it et al.}, Physica C {\bf 263} 1996
(1996).

\bibitem{20} N.E. Hussey, {\it et al.}, Phys. Rev. B {\bf 57}, 5505
(1998).

\bibitem{21b} B.L. Brand, D.W. Liu and L.G. Rubin {\bf 70}, 104 (1999).

\bibitem{21} T. Oguchi, Phys. Rev. B {\bf 51}, 1385 (1995); D.J.
Singh, Phys. Rev. B {\bf 52}, 1358 (1195).

\bibitem{22} J.M. Ziman, {\it Principles of the Theory of Solids,
2nd Ed.}, (Cambridge Univ. Press, 1972).

\bibitem{23} P.T. Coleridge, J. Phys. F:  Met. Phys. {\bf 17}, L79
(1987).

\bibitem{24} G.R. Stewart, Rev. Mod. Phys. {\bf 56}, 755 (1984).

\bibitem{25} B. H. Brandow, Phys. Rev. B {\bf 33}, 215 (1986).

\bibitem{26} T. Imai {\it et al.}, Phys. Rev. Lett. {\bf 81}, 3006
(1998).

\bibitem{27} N. Shirakawa {\it et al.}, J. Phys. Soc. Jpn. {\bf 64},
1072 (1995).

\bibitem{28} A. P. Mackenzie {\it et al.}, Phys. Rev. B {\bf 54},
7425 (1996).

\bibitem{29} M. Miyazawa, H. Kontani, and K. Yamada, preprint.

\bibitem{30} H. Mukuda, {\it et al.}, J. Phys. Soc. Jpn. {\bf 67},
3945 (1998).

\bibitem{31} I.I. Mazin and D.J. Singh, Phys. Rev. Lett. {\bf 79},
733 (1997).

\bibitem{32} Y. Sidis {\it et al.}, preprint (cond-mat/9904348, 23 Apr.,
1999).

\bibitem{33} I.I. Muzin and D.J. Singh, preprint (cond-mat/9902193,
12 Feb., 1999).

\bibitem{33b} N. Shirakawa, {\it et al.}, Phys. Rev. B {\bf 56}, 7890
(1997).

\bibitem{33c} K. Yoshida, {\it et al.}, Phys. Rev. B {\bf 58}, 15062
(1998).

\bibitem{33d} C. Pfleiderer, {\it et al.}, Phys. Rev. B {\bf 55}, 8330
(1997).

\bibitem{33e} J.A. Hertz, Phys. Rev. B {\bf14}, 1165 (1976);  
	A.J. Millis, Phys. Rev. B {\bf 48}, 7183 (1993).

\bibitem{34} For a recent review, see T. Timusk and B. Statt, Rep.
Prog. Phys. {\bf 62}, 61 (1999).

\bibitem{35} N. W. Preyer {\it et al.}, Phys. Rev. B {\bf 44}, 407
(1991); A. Lacerda, {\it et al.}, Phys. Rev. B {\bf 49}, 9097
(1994); J. M. Harris, {\it et al.}, Phys. Rev. Lett. {\bf 75}, 1391
(1995); T. Kimura, {\it et al.}, Phys. Rev. B {\bf 53}, 8733 (1996).

\bibitem{36} B. Batlogg, {\it et al.}, Physica C {\bf 140}, 235
(1994).

\bibitem{36b} R. Jin, {\it et al.}, J. Phys. Chem. Solids {\bf 59}, 2215
(1998).

\end{references}
\end{document}